\documentclass[showpacs,preprintnumbers,amsmath,amssymb,floatfix]{revtex4}
\usepackage{graphicx}
\usepackage{amsfonts,amsmath,amssymb}
\usepackage{dcolumn}
\usepackage{bm}

\begin{document}
\def\bbox#1{\hbox{\boldmath${#1}$}}
\def\blambda{{\hbox{\boldmath $\lambda$}}}
\def\eeta{{\hbox{\boldmath $\eta$}}}
\def\bxi{{\hbox{\boldmath $\xi$}}}
\setlength{\topmargin}{0.05in}
\newlength{\figwidth}
\setlength{\figwidth}{1.0\textwidth}

\title{Peculiar Features of the Interaction Potential between Hydrogen
and Antihydrogen at Intermediate Separations}

\author{Teck-Ghee~Lee} \email[Corresponding author email:
]{leetg@ornl.gov}

\affiliation{Physics Division, Oak Ridge National Laboratory, Oak
Ridge, TN 37831}

\affiliation{Department of Physics and Astronomy, University of
Kentucky, Lexington, KY 40506}

\author{Cheuk-Yin Wong} \email{wongc@ornl.gov}

\affiliation{Physics Division, Oak Ridge National Laboratory, Oak
Ridge, TN 37831}
\author{Peter Lee-Shien Wang}

\affiliation{Physics Division, Oak Ridge National Laboratory, Oak
Ridge, TN 37831}

\affiliation{Department of Physics, Harvey Mudd College,
Claremont, CA 91711}

\date{\today}

\begin{abstract}
We evaluate the interaction potential between a hydrogen and an
antihydrogen using the second-order perturbation theory within the
framework of the four-body system in a separable two-body basis.  We
find that the $H$-$\bar H$ interaction potential possesses the
peculiar features of a shallow local minimum located around
interatomic separations of $r\sim$ 6 a.u. and a barrier rising at $r
\lesssim 5$ a.u.  Additional theoretical and experimental
investigations on the nature of these peculiar features will be of
great interest.
\end{abstract}


\maketitle

\section{Introduction}
Recent production of cold antihydrogen \cite{Amoretti,Gabrielse} has
stimulated new theoretical and experimental interests on the $H$ and
the $\bar H$ system. The four particles of the greater $(e^+e^-p\bar
p)$ system can be arranged in different ways leading to different
types of states.  On the one hand, there are states in the $H$-$\bar
H$ family in which the $e^+$ and the $e^-$ are not correlated but are
orbiting around the $\bar p$ and the $p$ respectively, with the $H$
and the $\bar H$ appearing as composite particles.  They are the
lowest-energy states at large interatomic separations.  On the other
hand, there are states of the $Psp\bar p$ family for which the $e^+$
and the $e^-$ are highly correlated and orbit each other as in a
positronium (Ps).  The role of the $Psp\bar p$ family and the
$H$-$\bar H$ family is interchanged at short distances as the
$e^+$-$e^-$-correlated state of the $Psp\bar p$ family becomes the
state of the lowest energy, as indicated by the merging of the
$V_{var}(r)$ potential obtained by variational calculations with the
$V_{Psp\bar p}(r)$ potential obtained for the $Psp\bar p$
configuration, at small interatomic separations (see Fig.\ 4 below).
Because the $p$ and $\bar p$ can orbit around (and annihilate) each
other under their mutual Coulomb and nuclear interactions
\cite{Won84}, the $Psp\bar p$ family can be further bifurcated into
subfamilies with a correlated or uncorrelated $p$-$\bar p$ pair. There
are also states that are hybrid of different families.

If one can depict the orbiting of one particle relative to another
particle as a ``dance pattern'', then the dance patterns of the
different families have distinctly different topological structures
and connectivities. The distinct topological structures may allow them
to retain some of their characteristics and stability.  The mixing and
the interplay between different families at different interatomic
separations will provide a general idea on the ``genealogy" of the
states.  To be able to classify the greater $(e^+e^-p\bar p)$ system
into families, if at all possible, will bring us to a better
understanding of the complexity of the spectrum that is associated
with the complicated four-body problem, as well as to a better
knowledge of the scattering between the $H$ and the $\bar H$.

The $H$ atom and the $\bar H$ atom can scatter from each other at
various energies.  The problem of the $H$-$\bar H$ scattering will
require the knowledge of the interaction potential between the $H$ and
the $\bar H$ as composite particles.  As the $H$ and $\bar H$ may
rearrange into other configurations and may annihilate at short
distances, the scattering problem needs to be treated in general in a
coupled-channel analysis involving different interaction potentials
and transition matrix elements \cite{Zyg04}.

For the elastic channel, one often takes the interaction potential to
be the adiabatic potential obtained in a variational calculation for
the lowest-energy state of the $e^+$ and the $e^-$ in the
Born-Oppenheimer potential
\cite{Zyg04,Kolos,Armour,Froelich,Strasburger,Labzowsky,Sharipov}.
The $e^+$-$e^-$ correlation property of the lowest-energy adiabatic
state will undergo a change as the interatomic separation changes.  At
large interatomic separations, the adiabatic lowest-energy state
coincides with the incoming channel state of separated $H$ and $\bar H$
atoms with uncorrelated $e^-$ and $e^+$ orbiting around $p$ and $\bar
p$, respectively.  At small interatomic separations, the variational
calculations will naturally lead to the lowest-energy adiabatic state
involving strong correlations between the $e^+$ and the $e^-$ which
differs from the elastic incoming channel by the presence of
$e^+$-$e^-$ correlations.  An adiabatic approximation involves
altering the nature of the underlying state from the incoming elastic
channel with no $e^+$-$e^-$ correlations to the other
$e^+$-$e^-$-correlated channel, as the interatomic separation
decreases.  A general consideration of the channel coupling however
will contain both the incoming elastic channel and the
$e^+$-$e^-$-correlated channel in explicit coupling and
transition. Depending on the collision energy, the dynamics of the
scattering process will lead to the adiabatic case with strong
$e^+$-$e^-$ correlations at short separations in one limit of slow
motion and large transition matrix elements.  It will lead to the
diabatic case with no $e^+$-$e^-$ correlation of the atomic $H$ and
$\bar H$ in the other limit.  To carry out the explicit channel
coupling for the general case of an arbitrary energy, or to see the
cross-over to either the adiabatic or the diabatic limit, it is
necessary to obtain the interaction potential for the incoming
entrance channel of a $H$ and an $\bar H$.

In the present work, we focus our attention on the interaction
potential for this incoming entrance channel between the $H$ and
the $\bar H$ as composite particles, for which the $e^-$ and the
$e^+$ remain uncorrelated, orbiting around the $p$ and $\bar p$
respectively, i.e., without the loss of the $H$ and the $\bar H$
atomic identities. We are in effect trying to construct the
interaction potential for one of the diabatic basis configurations
in preparation for a general coupling of different diabatic basis
states in a couple-channel analysis.  For brevity of nomenclature,
we shall use the term ``the interaction potential'' to refer to
this potential between the composite particles $H$ and $\bar H$,
unless otherwise stated.

The interaction potential $V_{H\bar H}(r)$ between the composite $H$
and the $\bar H$ can also be used to investigate possible $H$-$\bar H$
two-body molecular states of the $H$-$\bar H$ family, as the composite
$H$ and $\bar H$ atoms can form a bound system if their interaction
potential is sufficiently attractive.  These possible molecular
two-body $H$-$\bar H$ states, if they exists, belongs to the $H$-$\bar
H$ family, a subset of the states of the greater four-body
$(e^+e^-p\bar p)$ system.  One theoretical approach to study the
lowest-energy state of the greater $(e^+e^-p\bar p)$ system is to use
the variational method equipped with either explicitly correlated
Slater-type or Gaussian-type orbitals with $e^+$-$e^-$
correlations. This method has provided a great wealth of information
regarding the lowest-energy state of the four-body system
\cite{Armour,Kolos,Froelich,Strasburger,Labzowsky}.  As our interest
is focused on the properties of possible two-body states of the
composite $H$ and $\bar H$ atoms, the two-body $H$-$\bar H$ states of
our interest may or may not necessarily be the lowest-energy state of
the greater $(e^+e^-p\bar p)$ system.  Its relative position depends
on the interatomic separation.  We mentioned earlier that at large
interatomic separations, the two-body state with separated atoms and
leptons in the $1s$ orbitals is the lowest-energy state of the greater
$(e^+e^-p\bar p)$ four-body system.  At small interatomic separations,
the two-body molecular state of the $H$ and $\bar H$ lies higher than
the variational adiabatic lowest-energy state which is characterized
by strong $e^+$-$e^-$ correlations, and the two-body molecular state
of the $H$ and $\bar H$ is an excited state of the greater
$(e^+e^-p\bar p)$ system. It will be of great interest in future work
to investigate how the two-body $H$-$\bar H$ state may preserve or
otherwise mix their distinct molecular characteristics with the
$e^+$-$e^-$-correlated state, as the system dynamically traverses from
large interatomic separations to the region of small separations.

To follow the four-body problem in the two-body atomic basis, we
study the system from {\it outside-in} as the $H$ and the $\bar H$
approach each other, and we examine their virtual excitations into
excited $H$-$\bar H$ configurations, as the interatomic distance
decreases.  A good method to carry out such an investigation is to
use the perturbation theory, having the atomic $H$ and $\bar H$ as
unperturbed composite particles, as in previous study of molecular
states in heavy mesons in hadron physics \cite{Won04}. The
perturbation theory can be justified as a useful tool here because
the composite $H$ and $\bar H$ atoms are neutral objects, and
their residue interaction $V_I$ between the $H$ and the $\bar H$
contains four terms (as given in Eq.\ (\ref{VI}) below) involving
many cancellations due to the opposite charges of $e^-$ and $p$ of
one atom on the one hand, and additional cancellations due to the
opposite charges of the $e^+$ and $\bar p$ of the other antiatom
on the other hand.  As a consequence, the residue interaction is
small in magnitude.  Using ${V_I}/{(\Delta E)}$, where $V_I =
{\bbox{r_a}\cdot\bbox{r_b}}/{ r^3}$ is the dipole-dipole
interaction and $\Delta E$ is the energy denominator from the
ground to the first excited state of the $H$ atom, we roughly
estimated ${V_I}/{\bigtriangleup E}$ at interatomic separation $r$
= 3, 5 and 8 a.u. to be $\sim$ 0.05, 0.01 and 0.003, respectively.
Clearly, the next higher-order term involving the ratio
$({V_I}/{\bigtriangleup E})^2$ will be even smaller and the
perturbation series is expected to converge. Therefore these
numbers justify the application of the perturbation theory to the
$H$-$\bar H$ interaction potential in intermediate interatomic
distances (i.e., 3 $\leq r \leq$ 10 a.u.).

Within the perturbation theory, the $H$-$\bar H$ interaction potential
is at present known only within the leading order \cite{Mor73}. It is
positive for large distances, with a barrier whose peak lies in the
region of short distances.  It decreases precipitously at very short
distances. Previous analysis of the next-to-leading order interaction
potential between the $H$ and the $\bar H$ assumed an expansion of the
energy denominator up to the first order of the state energies
\cite{Mor73}. Such an expansion may not be accurate.  It is of
interest to re-evaluate more accurately this interaction potential,
$V_{H\bar H}(r)$, up to the next-to-leading order (second order) in
the residue interaction $V_I$. We shall be interested in the
interaction potential for the elastic channel for which the initial
incoming and the final outgoing $H$ and $\bar H$ atoms are in their
respective ground states at asymptotic separations.  Interaction
potential for states with excited $H$ and $\bar H$ atoms at asymptotic
separations can be similarly considered in a simple generalization.

For small interatomic distances, the $e^{+}$-$e^{-}$ correlations
become considerably important and we shall also study this additional
$e^{+}$-$e^{-}$ correlation effects on the potential energy around
this region of separation. Needless to say, it is necessary in the
future to include a greater number of the degrees of freedom, to take
into account addition types of polarization of the two-body
states. The perturbation theory results obtained here can provide a
benchmark against which additional distortions may be measured. They
can bring us to a better understanding of the scattering potential and
the complexity of the spectrum that is associated with the four-body
system.

This paper is organized as follows. In Section II, we review the
formulation of the four-body problem in terms of the interaction of
two composite objects using separable two-body basis. In Section III,
we show how to evaluate the interaction matrix elements and the
interaction potential. The results of the $H$-$\bar H$ interaction
potential and the examination of the stability with respect to
$e^+$-$e^-$ correlations are given in Sections IV and V,
respectively. Finally, Section VI gives some discussions and
conclusions of the present work. Some of the details of the analytical
formulas are presented in the Appendix.  Atomic units
$m$=$\hbar$=$e$=1 are used throughout the paper unless otherwise
indicated.

\section{The Four-body Problem in a Separable Two-Body Basis}

\begin{figure} [h]
\includegraphics[scale=0.45]{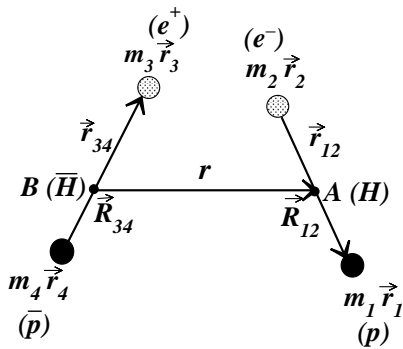}
\caption{ The coordinates of the $\{p,e^-,e^+,$ and $\bar p \}$
system.} \label{hhbar1234}
\end{figure}

We shall review the formulation of the four-body problem in a
separable two-body basis in terms of the interaction of two composite
particles as presented previously \cite{Won04}. As applied to the
$H$-$\bar H$ system, we choose the four-body coordinate system as shown
in Fig.\ \ref{hhbar1234} and label constituents $p$, $e^{-}$,
$e^{+}$, and $\bar p$ as particles 1, 2, 3, and 4, respectively with a
non-relativistic Hamiltonian
\begin{eqnarray}
\label{eq1} H=\sum_{j=1}^4 \frac{ \bbox{p}_{j}^2} { 2 m_j} +
\sum_{j=1}^4 \sum_ {k>j}^4 V_{jk} +\sum_{j=1}^4 m_j,
\end{eqnarray}
in which particle $j$ has a momentum $\bbox{p}_j$ and a rest mass
$m_j$.  The pairwise interaction $V_{jk}(\bbox{r}_{jk})$ between
particle $j$ and particle $k$ depends on the relative coordinate
between them, $\bbox{r}_{jk}=\bbox{r}_j-\bbox{r}_k.$
We introduce the two-body momentum $ {\bbox P}_{jk}=\bbox
{p}_j+\bbox{p}_k, $ and the two-body internal relative momentum $
\bbox{p}_{jk}=f_k \bbox{p}_j-f_j \bbox{p}_k, $ where $f_k=m_k/m_{jk}$,
$m_{jk}=m_j+m_k$.  We partition the Hamiltonian so that in the lowest
order the state in question can be described by a state of the
unperturbed Hamiltonian. We choose to partition $H$ into an
unperturbed Hamiltonian $h_{12}+h_{34}$ of atoms $A(12)$ and $B(34)$
according to
\begin{eqnarray}
H=\frac{ \bbox{P}_{12}^2}{2 m_{12}} + \frac{ \bbox{P}_{34}^2}{2
m_{34}} +V_I + h_{12} + h_{34} ,
\end{eqnarray}
\begin{eqnarray}
\label{VI}
V_I=V_{14}(\bbox{r}_{14})+V_{13}(\bbox{r}_{13}) +
V_{23}(\bbox{r}_{23})+V_{24}(\bbox{r}_{24}),
\end{eqnarray}
\begin{eqnarray}
h_{jk}=\frac{ \bbox{p}_{jk}^2}{2 \mu_{jk}} +
V_{jk}(\bbox{r}_j-\bbox{r}_k) + m_{jk} {\rm
~~~~for~~}(jk)=A(12){\rm~and~}B(34),
\end{eqnarray}
where $\mu_{jk}=m_j m_k/m_{jk}$.  The eigenvalues of the
Hamiltonians $h_{12}$ and $h_{34}$ can be solved separately to
obtain the bound state wave functions and masses $M_{jk}(\nu)$ of
atoms $A$ and $B$,
\begin{eqnarray}
h_{jk}|(jk)_\nu\rangle = [ \epsilon_{jk}(\nu) + m_{jk}]
|(jk)_\nu\rangle =M_{jk}(\nu)  |(jk)_\nu\rangle.
\end{eqnarray}
The four-body Hamiltonian becomes
\begin{eqnarray}
H=\frac{ \bbox{P}_{12}^2}{2 m_{12}} + \frac{ \bbox{P}_{34}^2}{2
m_{34}} +V_I + M_{12}(\nu)+ M_{34}  (\nu') .
\end{eqnarray}
The above non-relativistic approximation is obtained from
relativistic results by neglecting terms of order
$\epsilon_{jk}/M_{jk}$ and higher \cite{Won01}.  In order to
satisfy the boundary condition at large separations for which
$V_I$ approaches zero, we need to include some of these
higher-order terms and modify $m_{12}$ and $m_{34}$ in the above
equation to $M_{12}(\nu)$ and $M_{34}(\nu')$ so that the
Hamiltonian becomes
\begin{eqnarray}
H=\frac{ \bbox{P}_{12}^2}{2 M_{12}(\nu)} + \frac{
\bbox{P}_{34}^2}{2 M_{34}(\nu')} +V_I + M_{12}(\nu)+ M_{34}
(\nu').
\end{eqnarray}
The above Hamiltonian then describes properly the asymptotic
behavior at large separations of the two atoms.
This Hamiltonian ${H}$ is related to the standard
 Hamiltonian
\begin{eqnarray}
H_{\rm stand}= \frac{ \bbox{p}_{e}^2} { 2 m_e} + \frac{
\bbox{p}_{e^+}^2} { 2 m_{e^+}} + \sum_{j=1}^4 \sum_ {k>j}^4 V_{jk}
+\sum_{j=1}^4 m_j, \label{eq4}
\end{eqnarray}
by $ H=H_{\rm stand} +
\epsilon_{12}(\nu)+\epsilon_{34}(\nu')+\Delta, $
\begin{eqnarray}
\Delta&=&
 \frac{ \bbox{p}_{p}^2} { 2 m_p}
+ \frac{ \bbox{p}_{\bar p}^2} { 2 m_{\bar p}} +
\bbox{P}_{12}^2\left ( \frac{ 1}{2 M_{12}(\nu)} -\frac{ 1}{2
m_{12}} \right )
+ \bbox{P}_{34}^2\left ( \frac{ 1}{2 M_{34}(\nu')} -\frac{ 1}{2
m_{34}} \right ), \label{eq5}
\end{eqnarray}
and $\Delta$ is a small quantity which can be neglected in the
Born-Oppenheimer limit of fixed baryon centers.

We consider a four-body system in which both atoms will be in their
ground states $|A_0(12) B_{0}(34) \rangle$ at asymptotically large
interatomic separations where the residue interactions $V_I$
vanishes. When we include $V_I$ as a perturbation, the eigenfunction
of ${H}$ becomes
\begin{eqnarray}
\label{wave}
&\Psi&\!\!\!(\bbox{r},\bbox{r}_{12},\bbox{r}_{34})
=\psi(\bbox{r})\Biggl \{|A_0 B_{0}\rangle
+ {\sum_{\lambda,\lambda'}}' \frac{|A_{\lambda}
B_{\lambda'}\rangle \langle A_{\lambda} B_{\lambda'} | V_I | A_0
B_{0} \rangle} {\epsilon_A({\lambda})+\epsilon_B({\lambda'})-
\epsilon_A(0)-\epsilon_B(0)} \Biggr \},
\label{eq6}
\end{eqnarray}
where $\bbox{r}= \bbox{R}_{12}-\bbox{R}_{34}$ is the interatomic
separation (see Fig.\ 1), $\bbox{R}_{jk} = f_j~\bbox{r}_{j}+
f_k~\bbox{r}_{k}$ is the center-of-mass coordinate of $m_j$ and $m_k$,
and $\sum_{\lambda \lambda'}'$ indicates that the sum is over all
atomic states, including continuum states, except $| A_{0}
B_{0}\rangle$.  The eigenvalue equation is
\begin{eqnarray}
\label{sch}
H\Psi (\bbox{r},\bbox{r}_{12},\bbox{r}_{34})
=[M_{12}(0)+M_{34}(0)+\epsilon]
 \Psi (\bbox{r},\bbox{r}_{12},\bbox{r}_{34}).
\end{eqnarray}
Working in the center-of-mass frame and taking the scalar product of
the above equation with $|A_{0} B_{0}\rangle$, we obtain the
Schr\"odinger equation for the motion of $A_{0}(12)$ relative to
$B_{0}(34)$,
\begin{eqnarray}
\label{sch1} \left \{\frac{\bbox{p}^2}{2 \mu}_{AB} + V(\bbox{r})
\right \} \psi(\bbox{r}) =\epsilon \psi(\bbox{r}),
\end{eqnarray}
where $\bbox{p}$ is the relative momentum of the composite particles
\begin{eqnarray}
\bbox{p}=
\frac {M_{34}(0)\bbox{P}_{12}-M_{12}(0)\bbox{P}_{34}}
      {M_{12}(0)+ M_{34}(0)},
\end{eqnarray}
and $\mu_{AB}$ is the reduced mass of the two atoms
\begin{eqnarray}
\mu_{AB}=\frac {M_{12}(0)M_{34}(0)}
          {M_{12}(0)+M_{34}(0)}.
\end{eqnarray}

\section{Evaluation of $H$ and $\bar H$ Interaction Matrix Elements}

The Schr\"odinger equation (\ref{sch1}) involves the interatomic
separation $\bbox{r}$ and describes the motion between two
composite atoms.  Because the mass of the proton is much greater
than the mass of the electron and the positron, the interatomic
separation $\bbox{r}$ is approximately equal to the internuclear
separation $\bbox{r_{14}}$. The atom-atom potential $V(\bbox{r})$
in Eq.\ (\ref{sch1}) is given by
\begin{eqnarray}
\label{VVV} V(\bbox{r})&=&\langle A_0 B_{0} | V_I | A_0 B_{0}
\rangle -{\sum}_{\lambda,\lambda'}'\frac{ |\langle A_{\lambda}
B_{\lambda'} | V_I | A_0 B_{0} \rangle|^2}
{\epsilon_A({\lambda})+\epsilon_B({\lambda'})
-\epsilon_A({0})-\epsilon_B({0})},
\end{eqnarray}
where $|A_{0} B_{0}\rangle$ is a product of two 1s orbitals of the
hydrogen and the antihydrogen. We label $V(\bbox{r})$ as $V_{H \bar
H}(r)$, and introduce $E(\bbox{r})=V(\bbox{r})-1$.  We call the first
(leading-order) term on the right hand side the direct potential,
$V_{dir}(\bbox{r})$,
\begin{eqnarray}
\label{vdir0}
V_{dir}(\bbox{r})=\langle A_0 B_{0} | V_I | A_0 B_{0} \rangle,
\end{eqnarray}
and the second
(next-to-leading order) term the polarization potential,
$V_{pol}(\bbox{r})$,
\begin{eqnarray}
\label{vpoleq}
V_{pol}(\bbox{r})=-{\sum}_{\lambda,\lambda'}'\frac{ |\langle A_{\lambda}
B_{\lambda'} | V_I | A_0 B_{0} \rangle|^2}
{\epsilon_A({\lambda})+\epsilon_B({\lambda'})
-\epsilon_A({0})-\epsilon_B({0})}.
\end{eqnarray}

The direct potential is given by the sum of the four matrix elements
$\langle A_0 B_0| V_{jk}(\bbox{r}_{jk}) | A_0 B_0\rangle$ of Eq.\
(\ref{VI}).  From the $1s$ wave functions we can analytically
determine \cite{Mor73}
\begin{eqnarray}
\label{vdir}
\langle A_0 B_0| V_I | A_0 B_0\rangle = \frac{e^{-2r}}{r} \left
(-1-\frac{5}{8}r+\frac{3}{4}r^2+\frac{1}{6}r^3\right).
\label{eq16}
\end{eqnarray}

The polarization potential $V_{ pol}(r)$ is obtained as a double
summation of $\lambda$ and $\lambda'$, representing the virtual
excitation of the $H$ and $\bar H$ from 1s states to excited
$\lambda$-$\lambda'$ atomic states. It contains contributions where
$\lambda$-$\lambda'$ are bound-bound, bound-continuum (or vice-versa),
or continuum-continuum states.

For the evaluation of the bound-bound component of the polarization
potential, $V_{pol}^{bb}(r)$, we shall use the Fourier transform
method \cite{Sat83,Won04}.  We represent the internal wave functions in
$A_\lambda (12)$ and $B_{\lambda'}(34)$ by normalized eigenfunctions
of the hydrogen atom $\phi_{\lambda}^{A}(\bbox{r}_{12})$ and
$\phi_{\lambda'}^{B}(\bbox{r}_{34})$, respectively.  The matrix
elements $\langle A_\lambda B_{\lambda'}| V_I | A_0 B_{0}\rangle$ is
then a sum of four terms with $V_I$ given by Eq. (\ref{VI}).  Each of
these terms is given by
\begin{eqnarray}
\langle A_\lambda B_{\lambda'}| V_{jk}(\bbox{r}_{jk}) | A_0
B_{0}\rangle = \int \frac{d\bbox{p}}{(2\pi)^3} e^{i \bbox{p}\cdot
\bbox{r}} {\tilde \rho}_{\lambda 0}^A[f_A(jk)\bbox{p}] {\tilde
\rho}_{\lambda' 0}^B[f_B(jk)\bbox{p}] {\tilde v}_{jk}(\bbox{p}),
\label{eq8}
\end{eqnarray}
where
\begin{eqnarray}
{\tilde \rho}_{\lambda 0}^{A,B}(\bbox{p}) =\int {d\bbox{y}} e^{i
\bbox{p}\cdot \bbox{y}} \rho_{\lambda 0}^{A,B}(\bbox{y}),
\label{eq9}
\end{eqnarray}
and
\begin{eqnarray}
{\tilde v}_{jk}(\bbox{p}) =\int {d\bbox{r}_{jk}} e^{-i
\bbox{p}\cdot \bbox{r}_{jk}}
 V_{jk} (\bbox{r}_{jk}), \label{eq10}
\end{eqnarray}
with $\rho_{\lambda
0}^A(\bbox{r}_{12})$=$\phi_{\lambda}^*(\bbox{r}_{12})
\phi_0(\bbox{r}_{12})$, $\rho_{\lambda'
0}^B(\bbox{r}_{34})$=$\phi_{\lambda'}^*(\bbox{r}_{34})
\phi_0(\bbox{r}_{34})$, and $f_{\{A,B\}}(jk)$ are the coefficients of
the linear relation
\begin{eqnarray}
{\bbox{r}}_{jk}={\bbox{r}}+f_A(jk)~{\bbox{r}}_{12}+f_B(jk)~{\bbox{r}}_{34},
\end{eqnarray}
where the values of $f_{\{A,B\}}(jk)$ are given in Ref.\
\cite{Won04}.

To calculate the matrix element (\ref{eq8}), we fix the vector
$\bbox{r}$ to lie along the $z$-axis, and quantize the azimuthal
component of the magnetic quantum number $m$ to be projections along
the $z$-axis. We limit our consideration to intermediate states with
$l$=1 orbital angular momentum, as the dominant polarizing interaction
is the dipole-dipole excitation.  The bound-bound intermediate states
are then ($\lambda,\lambda'$)=($\{n,(l$=$1),m\},\{n',(l$=$1),m'\}$)
where $n$ and $n'$ are principal quantum numbers and $n,n' \geq 2$.
The matrix element $\langle A_\lambda B_{\lambda'}|
V_{jk}(\bbox{r}_{jk}) | A_0 B_{0}\rangle$ is non-vanishing only for
$\{\lambda=1m; \lambda'=1~(-m)\}$ with $\{m=-1,0,1\}$, when both atoms
are excited.

We write down the $1s$ and $np$ hydrogen wave functions explicitly and
obtain the Fourier transform of the transition density ${\tilde
\rho}_{\lambda 0}^{A,B}(\bbox{p})$ (see Eq.(A9) in Appendix A)
\begin{eqnarray}
\label{eq22}
{\tilde{\rho}}_{\lambda0}^{A,B}(\bbox{p})
&=&\frac{(2\pi)^{3/2}N_{00}N_{n1}Y_{00}(\hat{r})}{(n+1)^2}\sum_{k=0}^{n-2}
\left(
\begin{array}{c}
n+1\\
k
\end{array}
\right)\frac{\beta^{(n-2-k)}}{N'_{n1k}}(1-\beta)^k
F_{n1k}\left(\frac{\bbox{p}}{n+1}\right),
\end{eqnarray}
where $\beta=1/(n+1)$,
\begin{eqnarray}
\label{eq23}
N'_{n1k} =
\sqrt{\left[\frac{2(n+1)}{n}\right]^3\frac{(n-2-k)!}{2n(n+1)!}},
\end{eqnarray}
\begin{eqnarray}
\label{eq24}
F_{n1k}\left(\frac{\bbox{p}}{n+1}\right)=i~\sqrt{\frac{2}{\pi}\frac{(n-2-k)!}{(n+1)!}}
\frac{2^4~n^2}{\sqrt{n+1}}\left(\frac{\frac{np}{n+1}}{[(\frac{np}{n+1})^2+1]^3}\right)
\mathcal{C}^2_{n-2-k}\left(\frac{(\frac{np}{n+1})^2-1}{(\frac{np}{n+1})^2+1}
\right)Y_{\lambda}(\hat{p}),
\end{eqnarray}
and $\mathcal{C}^q_{i}(q)$ denotes the Gegenbauer polynomials.

Upon substituting the above Fourier transform of the transition
density ${\tilde \rho}_{\lambda 0}^{A,B}(\bbox{p})$ into Eq.\
(\ref{eq8}), we can carry out the angular part of the
integral analytically. We recognize
\begin{eqnarray}
e^{i \bbox{p}\cdot \bbox{r}} = \sum\limits_{l=0}^{\infty}
(2l+1)i^lj_l(pr)P_l(cos\theta_p),
\end{eqnarray}
and we can easily carry out the angular integral by using
\begin{eqnarray}
\int\limits_{0}^{2\pi} d\phi \int\limits_{-1}^{1} d\eta ~e^{i
\bbox{p}\cdot \bbox{r}}~Y_{1m}(\theta_p)~ Y_{1-m}(\theta_p)
=\begin{cases}
  j_0(pr) -2j_2(pr)& ~~~({\rm for~}m=0), \\
  j_0(pr)+ j_2(pr) & ~~~({\rm for~}m=1), \\
 \end{cases}
\end{eqnarray}
where $j_{l}(x)$ and $P_l(z)$ are a spherical Bessel
function and Legendre polynomial of a degree of $l$, respectively.
The remaining one-dimensional integral in the momentum space can be
carried out numerically.

For the evaluation of the bound-continuum $V^{bc}_{pol}(r)$ and
continuum-continuum $V^{cc}_{pol}(r)$ components of the polarization
potential, we use the bound and continuum Coulomb wave functions in
the configuration space \cite{Abram} and perform a six-dimensional
numerical integration involving the residue interaction $V_I$ to
obtain $\langle A_\lambda B_{\lambda'}| V_I | A_0 B_{0}\rangle$. The
integrand is confined to a finite spatial region as the transition
densities $\rho_{\lambda 0}^A(\bbox{r}_{12})$ and $\rho_{\lambda
0}^B(\bbox{r}_{34})$ contain the product of the continuum wave
function and the ground state wave function.  The sum over the
continuum states can be carried out by discretizing the continuum
spectrum which leads to converging results without much difficulty, as
the magnitude of the matrix element decreases for continuum states
with large momenta.  We check our results against previous
calculations of the $C_6$ dispersion-energy coefficient at large
interatomic separations.  Good agreement of the $C_6$ coefficient then
allows us to proceed to explore the polarization potential at shorter
interatomic separations.

\section{The $H$ and $\bar H$ Interaction Potential}

\begin{figure} [h]
\includegraphics[scale=0.5]{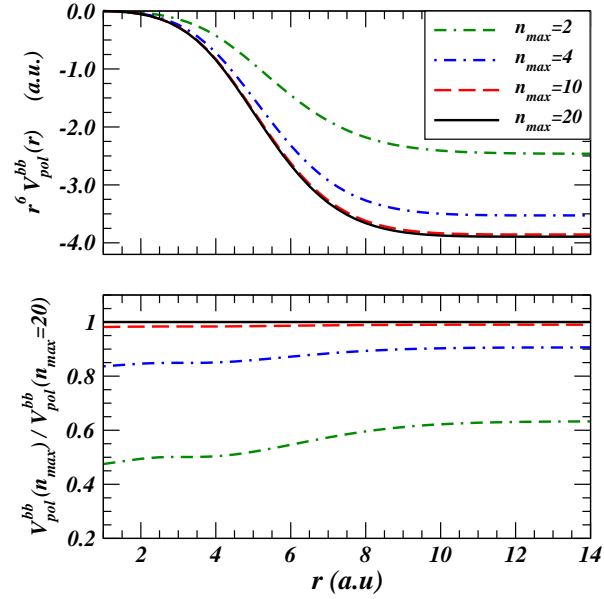} \caption{(a) Plot of
$V^{bb}_{pol}(r)\times r^6$ as a function of the interatomic
separation $r$ for different $n_{max}$ of $l=1$ states. (b) Ratio of
the summation from $2p$ up to $n_{\rm max}p$~states to the summation
from $2p$ up to $20p$ states as a function of the interatomic
separation $r$.}
\label{vpol}
\end{figure}

In the calculation to obtain the bound-bound contribution
$V_{pol}^{bb}(r)$ to the polarization potential, we include all $np$
states up to $n\leq n_{\rm max}$. Fig.~\ref{vpol} shows the
bound-bound results of $r^6 V^{bb}_{pol}(r)$ as a function of the
interatomic separation $r$ for different $n_{max}$ number of bound
states. Clearly, an increase in $n_{\rm max}$ for $n_{\rm max}>10$
brings essentially little change in the shape but an overall small
increase in the magnitude of the polarization potential, indicating
that the result is sufficiently converged by considering up to $n\leq
n_{\rm max}=20$ states.

\begin{figure} [h]
\includegraphics[scale=0.8]{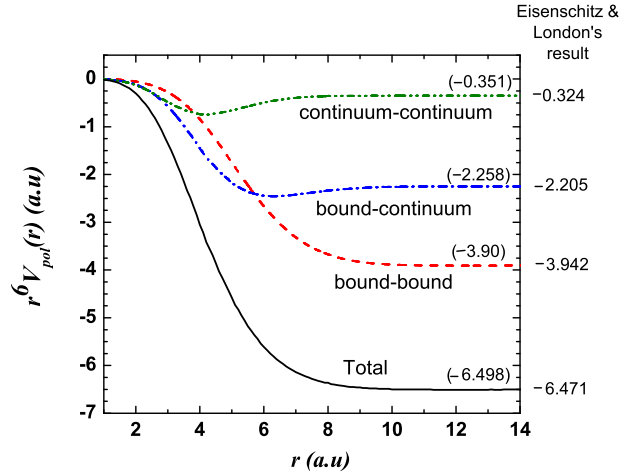}
\caption{Bound-bound, bound-continuum and continuum-continuum
components of the polarization potential multiplied by $r^6$ as a
function of the interatomic separation $r$.} \label{vpol_bb_bc_cc}
\end{figure}

Results for the bound-bound, bound-continuum and
continuum-continuum polarization potentials defined in Eq.
(\ref{vpoleq}) are presented in Fig.~\ref{vpol_bb_bc_cc} as a
function of the interatomic separation $r$. To exhibit the
asymptotic behavior of $V_{pol}\sim C_6 /r^6$ at large $r$, the
potentials are multiplied by $r^6$ in this figure.  The numerical
asymptotic values for each component are given in the parentheses,
to be compared with the values obtained by Eisenschitz and London
\cite{London}. While $r^6V_{pol}^{bb}(r)$ decreases monotonically
as a function of $r$, the bound-continuum $r^6V_{pol}^{bc}(r)$ and
continuum-continuum $r^6V_{pol}^{cc}(r)$ components have minima at
$r \sim$ 6 and $r \sim$ 4 a.u., respectively. The sum of all three
contributions leads to $r^6V_{pol}(r)$ decreasing monotonically as
a function of $r$. The result in Fig.\ 3 also illustrates that the
contribution of the continuum states to the polarization potential
is substantial and cannot be neglected.  The degree of the
modification when continuum intermediate states are included at
large $r$ are also known from the perturbation calculation for the
$C_6$ coefficient of H$_2$ by Eisenschitz and London
\cite{London}. Note that the best $C_6$ coefficient has been known
at least up to 8 digits of accuracy. Compare to the old
calculation \cite{London}, the present $C_6 = 6.498$ yields a
better agreement to the modern $C_6 = 6.4990267$ dispersion-energy
coefficient (i.e., less than $\sim$ 0.05\%).

\begin{figure} [h]
\includegraphics[scale=0.4]{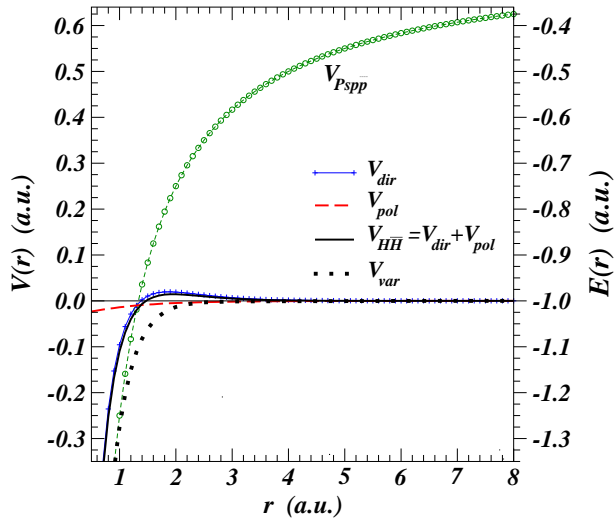}
\caption{The $H$-$\bar H$ interaction potential $V_{H\bar H}(r)$ and
its components $V_{dir}(r)$ and $V_{pol}(r)$, as a function of the
interatomic separation $r$.  Shown here are also the potential
$V_{Psp\bar p}(r)$ for the $ (e^+e^-)p \bar p$ configuration and the
potential $V_{var}(r)$ for the lowest-energy $ (e^+e^-p \bar p)$
four-body state obtained from the variational calculations of Ref.\
\cite{Strasburger}.}
\label{vhhbargros}
\end{figure}

The functional form of the polarization potential $V_{pol}(r)$ in
Fig.~\ref{vpol_bb_bc_cc} implies that while the polarization potential
varies as $C_6/r^6$ at large separations, they vary much more slowly
as a function of $r$ at small interatomic separations. This slow
variation and the relatively flat behavior can be seen from the total
polarization potential $V_{pol}(r)$ as a function of $r$ shown as the
dashed curve in Fig.~\ref{vhhbargros}.

\begin{figure} [h]
\includegraphics[scale=0.4]{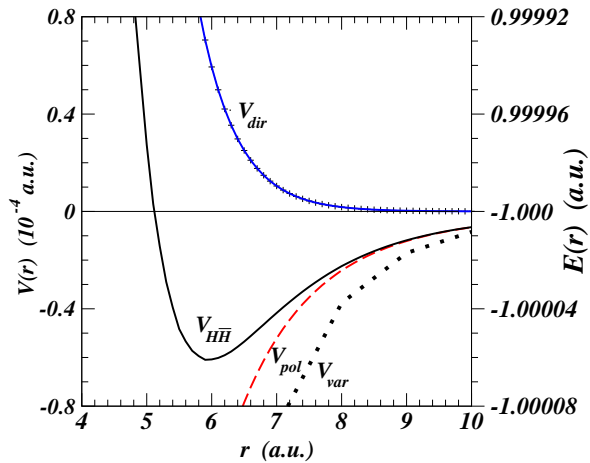}
\caption{A magnified view of Fig.\ 4 in the region of intermediate
separations. Shown here are the $H$-$\bar H$ interaction potential
of $V_{H\bar H}(r)$ and its components $V_{dir}(r)$ and
$V_{pol}(r)$, together with the potential $V_{var}(r)$
\cite{Strasburger}.}
\label{vhhbarfn}
\end{figure}

In the present calculation, we preserve the $H$ and $\bar H$ atomic
identities and let them interact through $V_I$ defined in
Eq.(\ref{VI}).  Within such a framework, the interatomic potential,
$V_{dir}$, $V_{pol}$ and the total $V_{H\bar H}$ are shown as various
curves in Figs.~\ref{vhhbargros} and \ref{vhhbarfn}.  In Fig. 4, the
direct potential $V_{dir}$ from Eq.~(\ref{vdir}) is slightly repulsive
at large distances with a barrier whose peak lies in the region of
small interatomic separations. It eventually decreases precipitously
at smaller interatomic separations. The polarization potential
$V_{pol}(r)$ is, however, always attractive (Fig.\ 3). The overall
magnitude of the polarization potential is small because the residue
interaction $V_I$ consists of four terms which tend to cancel among
themselves. The polarization potential provides a small overall
modification of the interaction potential at short distances (Fig.\ 4),
but it changes the character of the interaction potential at
intermediate separations of $r\sim 6$ a.u.\ where the direct potential
is small in magnitude, and the total interaction potential $V_{H\bar
H}(r)$ exhibits a shallow local minimum at $r \sim 6$ a.u.\ (Fig.\ 5).

To understand the rearrangement of the four particles in $H$ and
$\bar H$, we also plotted in Fig.~\ref{vhhbargros} the potential
$V_{var}(r)$ based on the variational calculation for the
lowest-energy state of the greater $(e^+e^-p\bar p)$ system
\cite{Strasburger}.  Asymptotically, the variational potential
$V_{var}(r\to \infty)$ goes to zero in energy and almost coincides
with $V_{H\bar H}(r)$ at large interatomic separations. At
intermediate interatomic separations, the variational potential
$V_{var}(r)$ lies below $V_{H\bar H}(r)$.  At very short distances
(i.e., $r \lesssim$ 2 a.u.),  $V_{var}(r)$ decreases
precipitously.

In Fig. 4, we also present the potential
\begin{eqnarray}
V_{Psp\bar p}(r) = -\frac{1}{r_{14}} - 0.25 + 1
\end{eqnarray}
for the $Psp\bar p$ configuration of a positronium and two heavy
nuclei.  Comparing $V_{var}(r)$ with $V_{Psp\bar p}(r)$ at large
$r$, the energy gap between them is 0.75 a.u. Only when $r$ is
small and the two curves start to meet at $r \sim 1$ a.u. will the
rearrangement from $H$-$\bar H$ to $ Psp\bar p$ becomes more
probable.  In the vicinity of the ``critical crossing-distance
$r_c$'' between the $V_{var}(r)$ and the $V_{Psp\bar p}(r)$
curves, the Born-Oppenheimer or adiabatic approximation starts to
break down \cite{Labzowsky,Stra2}.  Nevertheless, in order to
rearrange $H$-$\bar H$ to become $Psp\bar p$ as $r$ decreases
certainly required some changes in the topology of the four-body
system. At $r$ slightly greater than 1 a.u., the wave function may
be regarded as a mixture of the two families of $H\bar H$ and
$Psp\bar p$.  As the separation decreases further below $r < 1 $
a.u., the $V_{Psp\bar p}(r)$ and $V_{var}(r)$ potential curves
merge, indicating that the $Psp\bar p$ with strong $e^+$-$e^-$
correlations is the state of the lowest energy. The mixing of the
two families at small interatomic separations and the strong
$e^+$-$e^-$ correlations for $r<1$ a.u. is in agreement with our
analysis of stability against the variation of the $e^+$-$e^-$
correlations in the next Section.

\vspace{2.0cm}

In Fig. \ref{vhhbarfn}, we show a magnified view of the
interaction potential of $H$-$\bar H$ for interatomic separations
from 4 to 10 a.u. In this region, the sum of the direct potential
and the polarization potential gives rise to the $H$-$\bar H$
interaction potential that possess a shallow local minimum located
around $r\sim$ 6 a.u. and a barrier rising at $r \lesssim 5$ a.u.
The depth of the well is found to be $6 \times 10^{-5}$ a.u. or
$1.63$ meV, which may hold a bound state.

It is easy to understand the formation of a potential ``pocket''
for $H$-$\bar H$.  The system has the long-range attractive
potential behaving as $\sim C_6/r^6$ at large $r$ due to the
polarization of the atoms as they approach each other.  It
experiences a repulsion at shorter distances arising from the
repulsion of the $e^-$ with the $\bar p$ and the $e^+$ with the
$p$, which becomes effective at $ 1.5 < r < 5$ a.u. The
combination of a long-range attraction and a shorter range
repulsion results in a potential pocket at $r\sim 6$ a.u.  The
interaction however turns to become attractive at smaller
distances of $r\lesssim 1.5$ a.u. due to the attraction of the
leptons and the baryons.

Upon comparing with the variation potential $V_{var}(r)$, we
observe that $V_{H\bar H}(r)$ almost coincides with $V_{var}(r)$
at the asymptotic region of large $r$. As $r$ decreases, $V_{H\bar
H}$ deviates from and lies above $V_{var}(r)$ of the lowest-energy
state, indicating that the state of the $H$-$\bar H$ family is an
excited state. At this point, we would like to point out that our
aim here is to question, when there are no leptonic correlations
in our trial wave functions, how the neutral $H$ and $\bar H$
interact at intermediate distances before they dissolve completely
into positronium and protonium at shorter distances. Fig.\
\ref{vhhbarfn} provides the answer to our question.

\section{Stability against variations in $e^+-e^-$ correlations}

Unlike the interaction between two hydrogen atoms, there is no
exchange symmetry between the two leptons in the $H$-$\bar H$ system.
However, the correlation between the electron and the positron remains
important in the electrostatic energy calculations. Therefore it is
intriguing to see, within the present approach, when the $e^+$-$e^-$
correlation can play an important role in the energy of the
four-particle system. It is also crucial to find out whether the local
minimum at $r \sim 6$ a.u.\ examined in the last section is stable
against $e^+$-$e^-$ correlation perturbations. We therefore construct
first a ``poor man'' variational wave function with the variational
parameter $\eta$ that allows for $e^+$-$e^-$ correlations,
\begin{eqnarray}
\label{cor}
\Psi(\bbox{r},\bbox{r}_{12},\bbox{r}_{34},\bbox{r}_{23})=N_{\eta}
\{\sqrt{(1-\eta^2)}A_0(\bbox{r}_{12})B_0(\bbox{r}_{34}) +\eta
C_0(\bbox{r}_{23})D_0(\bbox{r}_{14})\},
\end{eqnarray}
where $N_{\eta} $ denotes the normalization constant such that for a
fixed $r$
\begin{eqnarray}
\int d\bbox{r}_{12}
d\bbox{r}_{34}|\Psi(\bbox{r},\bbox{r}_{12},\bbox{r}_{34},\bbox{r}_{23})|^2
= 1
\end{eqnarray}
and
\begin{eqnarray}
\label{CD} \int d\bbox{r}_{12}
d\bbox{r}_{34}|C_0(\bbox{r}_{23})D_0(\bbox{r}_{14})|^2 = 1.
\end{eqnarray}
The $C_0(\bbox{r}_{23})$ is the ground-state positronium wave function
centered at the center of mass origin and $D_0(\bbox{r}_{14})$ is a
plane wave state in the continuum. We shall examine the case of
$D_0(\bbox{r}_{14})$ to be a state with zero momentum (for $p$ and
$\bar p$ to be at rest) which supplies a constant to normalize the
product wave function in Eq.\ (\ref{CD}). We numerically evaluate
$\langle \Psi| {H_{\rm stand}}(r)| \Psi\rangle$ = $E(r)$ as a
six-dimensional integral in the configuration space.

\begin{figure} [h]
\includegraphics[scale=0.40]{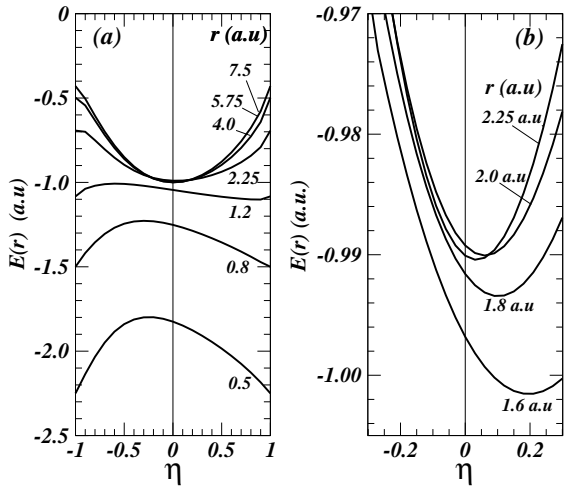}
\caption{$E$($r$) as a function  of $\eta$ and the interatomic
separation $r$. } \label{corrl}
\end{figure}

Figure \ref{corrl} gives the quantity $E(r)$ = $V_{H\bar H}$($r$)$-$1
as a function of the variation parameter $\eta$ for different
interatomic separations $r$.  For the largest values of $r$ in
Fig. \ref{corrl}($a$), the minimum of $E(r)$ resides at $\eta$ = 0,
indicating that the $H\bar H$ system is stable against $e^+$-$e^-$
correlations at large $r$. As $r$ decreases, the $E(r)$ potential
surface flattens as a function of $\eta$ and the $H$-$\bar H$ system
remains stable against $e^+$-$e^-$ correlations.  In the region where
the energy minimum begins to shift away from $\eta=0$, the variation
of the energy as a function of $\eta$ is shown for a finer increment
of $r$ in Fig.\ 6($b$).  At 2.25 $\ge r \ge$ 1.2 a.u., the minimum of
$E(r)$ is shifted to a positive value of $\eta$.  The equilibrium
configuration is one in which the state is a mixture of the $H\bar H$
and the $Psp\bar p$ configuration. The amplitude $\eta$ at equilibrium
increases as the interatomic separation decreases.  As a result, the
potential obtained from the variational method, $V_{ var}(r)$, is
lower than $V_{H\bar H}$ and $V_{Psp\bar p}$ in this range of
separation. At $r <$ 1.2 a.u., the lowest-energy state occurs at $\eta
= \pm 1$ (Fig.\ 6($a$)) which corresponds to a state with a completely
correlated $e^+$-$e^-$ component $C_0({\bbox{r}}_{23})D_0({\bf
r}_{14})$ and the absence of the uncorrelated component $A_0({\bf
r}_{12}) B_0({\bbox{r}}_{34})$.  The equilibrium configuration resides
in the $Psp\bar p$ configuration, indicating the dominance of
$e^+$-$e^-$ correlation at small $r$. This picture is consistent with
the findings of earlier variational calculations
\cite{Kolos,Armour,Froelich,Strasburger,Labzowsky} and
Fig.~\ref{vhhbargros}.

While Fig. \ref{corrl} gives the general feature of the potential
landscape with respect to $e^+$-$e^-$ correlations as $r$ changes, it
is necessary to investigate whether the local potential minimum at $r$
= 6 a.u.\ is stable against perturbations in $e^+$-$e^-$
correlations. We replace $|A_0B_0\rangle$ in Eq.\ (\ref{cor}) by the
wave function represented by the curly bracket of Eq.\ (\ref{wave}),
with the $|A_\lambda B_{\lambda'}\rangle$ amplitudes evaluated at $r$
= 6 a.u.\ and the summations over $\lambda$ and $\lambda'$ carried out
to include all $np$ states up to $n\leq n_{max}=20$.  We find that at
the location where $V_{H\bar H}(r)$ = $E(r)+1$ is a minimum in $r$,
$E(r)$ is also a minimum with respect to variations in $\eta$ at
$\eta=0$.  We conclude that the local potential minimum of $V_{H\bar
H}(r)$ at $r\sim 6$ a.u. is stable against perturbations from
$e^+$-$e^-$ correlations.

\section{Discussions and Conclusions}

The four-body system of $(e^+e^-p\bar p)$ contains a large degrees of
freedom as the particles can appear in many different combinations and
correlations, exemplified by the large number of parameters in an
elaborate variational calculation.  To be able to classify the states
of the greater $(e^+e^-p\bar p)$ system into simple families, if at
all possible, will bring us to a better understanding of the
complexity of the spectrum and the interactions that are associated
with the complicated four-body problem.

In terms of the correlation between the $e^+$ and the $e^-$, one
can perhaps roughly classify the greater $(e^+e^-p\bar p)$ system
as belonging to the $H$-$\bar H$ family without $e^+$-$e^-$
correlations and the $Psp\bar p$ family with strong $e^+$-$e^-$
correlations.  As one notes from energy considerations in Fig.
\ref{vhhbargros}, the lowest-energy state of the $(e^+e^-p\bar p)$
system is dominated by a state of the $H$-$\bar H$ family at
large interatomic separations but at short distances the system is
dominated by a state of the $Psp\bar p$ family.  The
lowest-energy state therefore changes its character as the
interatomic separation decreases.

In the collision of the $H$ and the $\bar H$, the four-body system
is prepared in the entrance channel of the $H$-$\bar H$ family as
the composite particles approach each other.  The system may
follow the configuration of the lowest energy adiabatically in a
slow collision, or it may follow the diabatic configuration of the
$H$-$\bar H$ entrance channel when the collision is fast or when
the intrinsic differences in the characteristics of the two
configurations hindered their transition.  In general, for the
collision of arbitrary energies, it is necessary to couple the
different channels explicitly and to follow the transition of the
channels as the composite particles approach each other.  We are
therefore motivated to examine the interaction potential between
the $H$ and the $\bar H$ as composite particles, for which the
$e^-$ and the $e^+$ remain uncorrelated, orbiting around the $p$
and $\bar p$ respectively.

We use the tool of the perturbation theory to examine this interaction
potential, with the states of $H$ and $\bar H$ as unperturbed basis
states.  The perturbation theory can be a useful concept because the
residue interaction $V_I$ contains many cancellations and is small in
magnitude compared to the value of typical energy denominators.  The
lowest-order perturbation theory gives an interaction potential
$V_{dir}(r)$ that is positive at large separations with a barrier
located approximately between 1.4 to 5.0 a.u.\ and a precipitous drop
at $r \lesssim$ 1.4 a.u.  Our calculations have been carried out up to
the next-to-leading order, by including virtual excitations of bound
and continuum $l=1$ states of the $H$ and $\bar H$ atoms.  The
next-to-leading order leads to the polarization potential that is
always attractive.  The bound-bound contribution to the polarization
potential has been obtained by using the Fourier transform method
involving the virtual excitation from the 1s states to all $np$ states
with $n \leq n_{\rm max}=20$. The bound-continuum, and the
continuum-continuum contributions have been obtained numerically by
discretizing the continuum. The polarization potential results agree
with previous calculations of the $C_6$ coefficient at large
separations.  The variation of the interatomic separation then allows
us to obtain the polarization potential $V_{pol}(r)$ at different
separations.

We find that the interaction potential between the composite $H$ and
the $\bar H$ possesses peculiar features.  At intermediate distances,
the total interaction potential $V_{H\bar H}(r)$, which is the sum of
the direct potential $V_{dir}(r)$ and the polarization potential
$V_{pol}(r)$, becomes negative.  At shorter distances, the total
potential shows a minimum at $r\sim 6$ a.u., a barrier rising at
$r\sim 5$ a.u., and a broad barrier peak at $r \sim 1.9 $ a.u., as
shown in Fig.~\ref{vhhbargros}.

Our investigation on the stability against variations in the
$e^+$-$e^-$ correlation in Section V indicates that the composite
system is stable against variations of the $e^+$-$e^-$ correlation for
$ r \gtrsim 3 $ a.u.\ but is unstable for $r \lesssim 3 $ a.u.  Thus,
the state with $e^+$-$e^-$ correlations is the state of lowest energy
at distances $r \lesssim 3$ a.u., but the composite $H$-$\bar H$ state
may remain stable without substantial $e^+$-$e^-$ correlations as an
excited system for $r \gtrsim 3 $ a.u.

The shallow local minimum located around interatomic separations
of $r\sim$ 6 a.u.\ has a depth of $6 \times 10^{-5}$ a.u.  One can
use the interaction potential $V_{H\bar H}(r)$ and solve the
Schr\"odigner equation (\ref{sch1}) for possible two-body
$H$-$\bar H$ states.  Even though the potential minimum is
shallow, a simple WKB calculation indicates that the potential is
deep enough to hold a bound state because of the large reduced
mass of the hydrogen and the antihydrogen.  The two-body $H$-$\bar
H$ state, if it exists, will not be the ground state, as it is
specialized to the configuration of the $H$-$\bar H$ family and
lies above the state of the lowest energy obtained in a variation
calculations.

It is necessary in the future to include a greater number of the
degrees of freedom, to take into account additional types of
polarization of the two-body states such as the virtual excitation of
the $d$, $f$,... states.  As the residue interaction is dominated by
the dipole-dipole interaction, the higher multipole interactions are
expected to be diminishing in their importance and will not modify
greatly the gross features of the interaction potential.  Because the
polarization potential is always negative, the addition of these
multipole interactions will probably only deepen slightly the local
potential minimum at $r\sim 6$ a.u.

Returning to the question of channel coupling and reaction processes,
we wish to emphasize that we have obtained only the interaction
potential $V_{H\bar H}$ for the channel of the $H$ and the $\bar H$ as
composite objects.  We have in effect constructed the interaction
potential only for one of the diabatic basis configurations.
Interaction potentials for other basis configurations can be
obtained in different analyses and a full picture of adiabatic or
diabatic motion for reactions at an arbitrary energy needs to be
investigated by a couple-channel analysis involving different diabatic
channels with different interaction potentials and transition matrix
elements.

In conclusion, we have examined the interaction potential of the
$H$-$\bar H$ system as composite objects using the second-order
perturbation theory with mutual polarizing excitations of the atoms.
We found that the interaction potential has a peculiar local minimum
at $r\sim 6$ a.u.  and a barrier rising at $r \lesssim 5$ a.u.  The
potential energy minimum is found to be stable against perturbations
of the $e^+$-$e^-$ correlation. Further definitive studies on the
nature of the interaction potential energy around this energy minimum
and the interplay between the $H$-$\bar H$ state and other states of
different configurations will be of great experimental and theoretical
interests.

\begin{acknowledgments}
The authors wishes to thank Prof. J.~H. Macek and Dr. S. Yu
Ovchinnikov and Prof. Cheng-Guang Bao for helpful discussions.  This
research is supported in part by the Division of Nuclear Physics, U.S.
D.O.E., under Contract No. DE-AC05-00OR22725, managed by UT-Battelle,
LLC.
\end{acknowledgments}

\appendix
\section{ Fourier transform of the transition density ${
\rho}_{\lambda0}^{A,B}(\bbox{r})$}

We here derive the expression for the Fourier transform of the
transition density, ${\tilde \rho}_{\lambda0}^{A,B}(\bbox{p})$, which
involves the product of $1s$ and $np$ wave functions. The hydrogen
wave function is given by
\begin{eqnarray}
\phi_{nlm}(\bbox{r}) =
N_{nl}\left(\frac{Z}{a_o}\right)^l\left(\frac{2r}{n}\right)^l
~\exp\left(\frac{-2Zr}{2na_o}\right)~\emph{L}^{2l+1}_{n-l-1}\left(\frac{2Zr}{na_o}\right)Y_{lm}(\hat{r}),
\end{eqnarray}
where $N_{nl}$ is the normalization constant, $Y_{lm}(\hat{r})$ is the
spherical harmonic and $\emph{L}^{(\alpha)}_{n}(x)$ is the Laguerre
polynomial. Setting $Z=1$, $l=$1 and $a_o=1$, we have
\begin{eqnarray}
{\rho}_{\lambda0}^{A,B}(\bbox{r})&&=\phi_{000}(\bbox{r})\phi_{n1m}(\bbox{r}) \nonumber \\
&&=N_{00}N_{n1}\left(\frac{2r}{n}\right)
~e^{-r(n+1)/n}
\emph{L}^{3}_{n-2}
\left(\frac{2r}{n}\right)Y_{00}(\hat{r})Y_{lm}(\hat{r}).
\end{eqnarray}
From Ref.\ \cite{Abram}, we have
\begin{eqnarray}
\emph{L}^{(\alpha)}_{j}\left(\beta x\right)= \sum_{k=0}^j \left(
\begin{array}{c}
j+\alpha\\
k
\end{array}
\right)\beta^{(j-k)}(1-\beta)^k~\emph{L}^{(\alpha)}_{j-k}(x).
\end{eqnarray}
Now by letting $r'=(n+1)r$ and writing
\begin{eqnarray}
f_{n1}(r')=\left(\frac{2r'}{n(n+1)}\right)~e^{-r'/n}
~\emph{L}^{3}_{n-2}\left(\frac{2r'}{n(n+1)}\right),
\end{eqnarray}
we use Eq.~(A3) with $j = n-2$, $\alpha=3$, and $\beta=1/(n+1)$ , and
we obtain
\begin{eqnarray}
f_{n1k}(r')=\sum_{k=0}^{n-2} \left(
\begin{array}{c}
n+1\\
k
\end{array}
\right)\beta^{(n-2-k)}(1-\beta)^k~\left(\frac{2r'}{n}\right)
~e^{-r'/n}
~\emph{L}^{3}_{n-2-k}\left(\frac{2r'}{n}\right).
\end{eqnarray}
Therefore, the transition density ${\rho}_{\lambda0}^{A,B}(\bbox{r'})$
becomes
\begin{eqnarray}
{\rho}_{\lambda0}^{A,B}(\bbox{r'})=\frac{N_{00}N_{n1}}{(n+1)}
f_{n1k}(r')Y_{00}(\hat{r})Y_{1m}(\hat{r}).
\end{eqnarray}
One can Fourier transform ${\rho}_{\lambda0}^{A,B}(\bbox{r'})$ by
\begin{eqnarray}
{\tilde{\rho}}_{\lambda0}^{A,B}(\bbox{p})
&=&\frac{(2\pi)^{3/2}N_{00}N_{n1}Y_{00}(\hat{r})}{(n+1)}
{\tilde{g}}_{\lambda0}^{A,B}(\bbox{p}),
\end{eqnarray}
where
\begin{eqnarray}
{\tilde{g}}_{\lambda0}^{A,B}(\bbox{p})&=&\frac{1}{(n+1)}\int
d\bbox{r'}
\exp \left \{ i \bbox{p} \cdot \bbox{r'} /(n+1)  \right \}
f_{n1}(r')Y_{lm}(\hat{r}) \nonumber \\
&=&\frac{1}{(n+1)}\int d\bbox{r'} e^{i \bbox{p'} \cdot \bbox{r'}}
N'_{n1k}f_{n1k}(r')Y_{1m}(\hat{r})/N'_{n1k},
\end{eqnarray}
with $\bbox{p'} = \bbox{p}/(n+1)$. By recognizing the latter
integrand is identical to the ``$p$ wave functions", without laborious
mathematics and algebra, one can make use of the formula for the
momentum space ``radial" wave function of hydrogen atom given in
Bransden and Joachain \cite{BJ} and cast the Fourier transform of the
above ${\rho}_{\lambda0}^{A,B}(\bbox{r'})$ into the results of Eqs.\
(\ref{eq22})-(\ref{eq24}).


\end{document}